\documentclass[onecolumn,floatfix,superscriptaddress,showpacs,showkeys,nofootinbib,preprint]{revtex4-1}

\usepackage{amsmath}
\usepackage{mathtools}
\usepackage[english]{babel}
\usepackage[bookmarks=false,colorlinks=false]{hyperref}
\usepackage{graphicx}
\usepackage{url}
 \usepackage{footnote}
 
\begin{document}

\title{The role of the local conservation laws in fluctuations of conserved charges}
%Trivial contributions to measured cumulants of net-particle fluctuations}

\author{P.~Braun-Munzinger}
\affiliation{Extreme Matter Institute EMMI, GSI, Darmstadt, Germany}
\affiliation{Physikalisches Institut, Universit\"{a}t Heidelberg, Heidelberg, Germany}
\author{A.~Rustamov}
\affiliation{Extreme Matter Institute EMMI, GSI, Darmstadt, Germany} \affiliation{Physikalisches Institut, Universit\"{a}t Heidelberg, Heidelberg, Germany}
\affiliation{National Nuclear Research Center, Baku, Azerbaijan}
\author{J.~Stachel}
\affiliation{Physikalisches Institut, Universit\"{a}t Heidelberg, Heidelberg, Germany}

\begin{abstract}
 In this report we present the first quantitative determination of the correlations between baryons and anti-baryons induced by local baryon number conservation. This is important in view of the many experimental studies aiming at probing the phase structure of strongly interacting matter. We confront our results with the recent measurements of net-proton fluctuations reported by the CERN ALICE experiment. The role of local baryon number conservation is found to be small on the level of second cumulants.
\end{abstract}
\maketitle

\section{Introduction} 
One of the key goals of nuclear collision experiments at high energy is to  map the phase diagram of strongly interacting matter.  A very challenging task is the determination of the QCD
phase structure including the search for critical behavior and verification of the possible existence of a critical end point of a first order phase transition line. A promising tool to probe the presence of critical behavior is the study of fluctuations of conserved charges since, in a thermal system,  fluctuations are directly related to the equation of state (EoS) of the system under the study.  One can  probe critical phenomena also at vanishing baryon chemical potential~\cite{LQCD1, Redlich1}. While at physical u,d quark masses a rapid cross over transition is found in Lattice QCD (LQCD), the proximity to a 2nd order phase transition of O(4) universality class suggests remainders of criticality. Indeed, the pseudo-critical temperature, reported from LQCD~\cite{LQCD1}, is in agreement with the chemical freeze-out  temperature as extracted by comparing  Hadron Resonance Gas (HRG) model predictions~\cite{HRG,Andronic:2018qqt} to the hadron multiplicities measured by ALICE. This agreement implies that strongly interacting matter, created in collisions of Pb nuclei at LHC energies, freezes out in close vicinity of the chiral phase transition line. Hence, singularities stemming from the proximity to a second order chiral phase transition can be captured also at vanishing net-baryon densities.
The current net baryon fluctuation measurements,  by the STAR collaboration at RHIC \cite{STARDATA,Xu:2018vnf}, and  by ALICE at the LHC \cite{RustamovQM17}, have provided interesting and stimulating results. However, quantitative analysis of these measurements  is made difficult by the presence of non-critical effects such as volume or participant fluctuations and by correlations introduced by  baryon number conservation. 
In statistical mechanics fluctuations of conserved quantities  are  predicted within the Grand Canonical Ensemble (GCE)~\cite{StatLandau} formulation, where  only the average values of these quantities are conserved~\cite{StatLandau}.  To compare theoretical calculations within a GCE, such as the HRG~\cite{HRG} and LQCD~\cite{LQCD1}, to experimental results, the requirements of GCE have to be achieved in experiments.  In experiments over the full acceptance, baryon number is conserved in each event, hence even in a limited acceptance its implications will be seen. Here, using the CE, we provide the first quantitative estimates of the implication of global and local baryon number conservation in a finite acceptance. 

\section{Fluctuations in GCE and CE}
In a thermal system with an ideal gas EoS, composed  of hadrons including baryon/anti-baryon species with baryon numbers +1 and -1, the GCE partition function  implies uncorrelated Poisson distributions for baryons and anti-baryons, hence the net-baryon distribution has the following cumulants~\cite{ourModel}\footnote{The probability distribution of the difference of two random variables each generated from uncorrelated Poisson distributions is called Skellam distribution.}: 

\begin{equation}
\kappa_{n}(Skellam)=\left<n_{B}\right>+(-1)^n\left<n_{\bar{B}}\right>,
\label{netcumulants}
\end{equation}
where  $\left<n_{B}\right>$ and $\left<n_{\bar{B}}\right>$ denote the first cumulants (mean numbers) of baryons and anti-baryons, respectively. 
Eq.~(\ref{netcumulants})  implies that ratios of even-to-even and odd-to-odd cumulants of net-baryons are always unity,
while the ratios of odd-to-even cumulants depend on mean multiplicities.
% or alternatively on $\mu/T$:

%\begin{alignat}{2}
\begin{equation}
\frac{\kappa_{2n+1}}{\kappa_{2k}}=\frac{\left<n_{B}\right>-\left<n_{\bar{B}}\right>}{\left<n_{B}\right>+\left<n_{\bar{B}}\right>}.
\label{conserv2}
%\end{alignat}
\end{equation}
\vspace{1mm}

Hitherto, the above conditions are used as baseline for net-baryon fluctuations. However, this can lead to misleading conclusions because, apart from dynamical fluctuations induced by critical phenomena, deviations from this baseline may be driven by non-dynamical contributions. Recently we demonstrated that fluctuations of participating nucleons from event-to-event significantly  distort measured event-by-event fluctuation signals. We also found that at LHC energies, where mean number of net-baryons measured at mid-rapidity is zero, contributions from participant fluctuations to the second and third cumulants of net-baryon distributions are vanishing~\cite{ourModel}.
\begin{figure}[htb]
\centering
 \includegraphics[width=0.45\linewidth,clip=true]{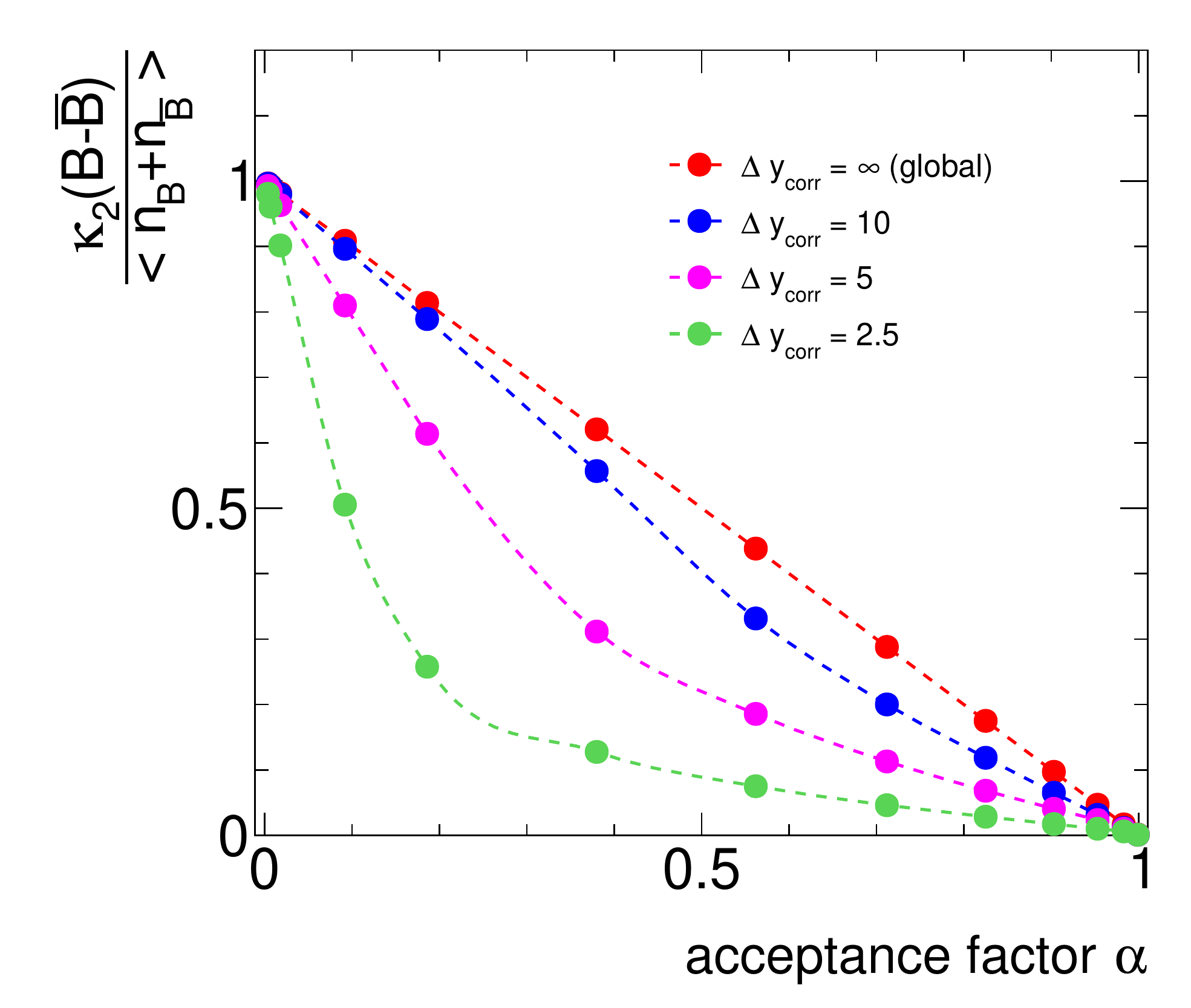}
 \includegraphics[width=0.45\linewidth,clip=true]{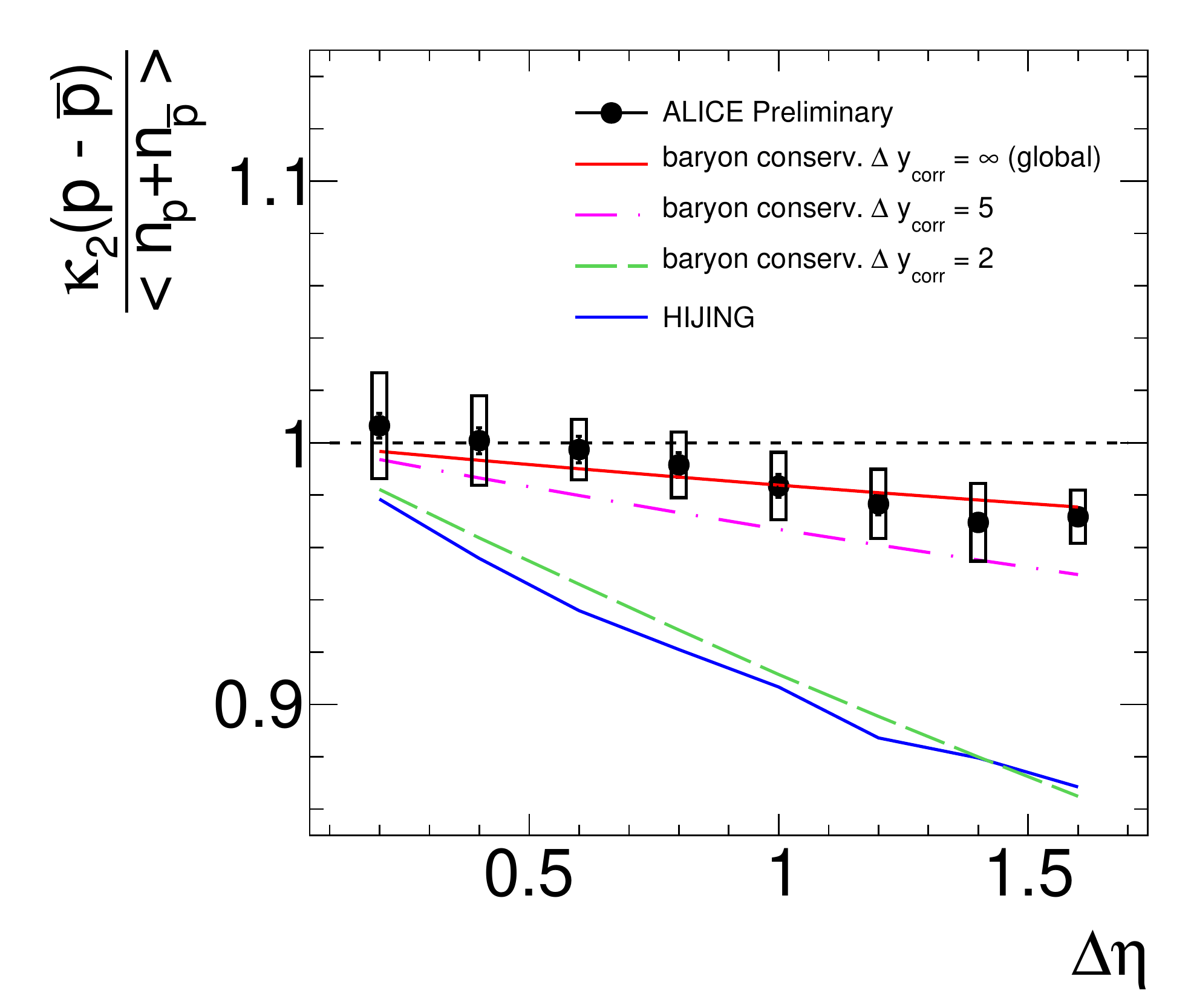} 
 \caption{Left panel: the normalized values of $\kappa_{2}(B-\bar{B})$, for different values of $\Delta y_{corr}$, as a function of accepted fraction of baryons. The red solid symbols, represented by $\Delta y_{corr} = \infty$, actually correspond to  $\Delta y_{corr}$ = 100, and are consistent with the global baryon number conservation. (cf. Eq.(4) of ~\cite{Braun-Munzinger:2018yru}). Right panel: comparison of the results with the ALICE data. Within the experimental uncertainties, the data are best described by global baryon number conservation ($\Delta y_{corr} = \infty$) but are consistent with $\Delta y_{corr} \ge 5.$  Values of  $\Delta y_{corr}$ smaller than 5 lead to results in disagreement with the experimental measurements. Interestingly, the blue solid line, representing the results of the HIJING generator, underestimates the experimental data and is described by the local baryon number conservation with $\Delta y_{corr}$ = 2.}
\label{fCE}
\end{figure} 
Below, we consider the CE partition function for a system of volume V, temperature T, and net-baryon number B to investigate effects of exact baryon number conservation. It is

\begin{equation}
Z_{CE}(V,T,B)=\sum_{N_{B}=0}^{\infty}\sum_{N_{\bar{B}}=0}^{\infty}\frac{(\lambda_{B}z_{B})^{N_{B}}}{N_{B}!}\frac{(\lambda_{\bar{B}}z_{\bar{B}})^{N_{\bar{B}}}}{N_{\bar{B}}!}\delta({N_{B}-N_{\bar{B}}-B})= \left(\frac{z_{B}}{z_{\bar{B}}}\right)^{\frac{B}{2}}I_{B}(2\sqrt{z_{B}z_{\bar{B}}})\bigg\rvert_{\lambda_{B,\bar{B}}=1},
\label{zCE}
\end{equation}
where $I_{B}$ denotes the modified Bessel function, $\lambda_{B,\bar{B}}$ are fugacities and $z_{B,\bar{B}}$ stand for single particle partition functions of baryons and anti-baryons respectively.
%$N_{B,\bar{B}}$ and $z_{B,\bar{B}}$ stand for GCE multiplicities and single particle partition functions of baryons and anti-baryons.
The $\delta$ function  in Eq.~(\ref{zCE}) guarantees that, in each event, the net number of baryons is fixed, i.e, net-baryons do not fluctuate from event-to-event.  In order to get finite fluctuations for net-baryons, distributions of baryons and anti-baryons have to be folded with the experimental acceptance. 
 
\section{Local Conservation Laws}
In~\cite{Braun-Munzinger:2018yru, koch-conserv, Schuster:2009jv, Keranen:2004eu}  effects of global conservation laws on fluctuations of conserved charges were addressed. In our previous work the energy dependence of cumulants of net-protons, reported by STAR for Au+Au collisions, is consistently described above $\sqrt{s_{NN}}$ = 11.5 GeV under the assumption of global baryon number conservation and fluctuations in the number of participating nucleons~\cite{Braun-Munzinger:2018yru}. Here, using  the same algorithm, we investigate contributions from local baryon number conservation~\cite{Vovchenko:2019kes}.  We first sample the number of baryons $N_{B}$ and anti-baryons $N_{\bar{B}}$  from the  probability distributions encoded in the CE partition function (cf. Eq.~\ref{zCE}). To this end, we simulated $10^{7}$ events with $\langle N_{B}\rangle$ = $\langle N_{\bar{B}}\rangle$ = 50 for baryons and anti-baryons respectively.\footnote{We verified that the presented results are not sensitive to the specific values of $\langle N_{B}\rangle$ and  $\langle N_{\bar{B}}\rangle$.} 
Next, using the shape of the charged particle pseudo-rapidity distribution as measured by ALICE~\cite{Abbas:2013bpa} and assuming that, at LHC energy, baryons follow the same shape as charged particles, we introduce finite acceptance effect. In doing so we first generate a baryon of rapidity
$y_{B}$, and a corresponding anti-baryon if its rapidity $y_{\bar{B}}$ satisfies the condition:
\begin{equation}
|y_{\bar{B}}-y_{B}|<\frac{\Delta y_{corr}}{2}.
\label{localY}
\end{equation}
We note that, in this representation,  global baryon number conservation corresponds to $\Delta y_{corr} = \infty$.  The results for the normalized values of $\kappa_{2}(B - \bar{B})$ are presented in the left panel of Fig.~\ref{fCE} as a function of the accepted fraction of baryons $\alpha$ for different values of $\Delta y_{corr}$. Here, $\alpha$ is defined as the ratio of baryons inside the acceptance to the number of baryons in full phase space. As expected, the magnitude of normalized net-baryon number fluctuations decreases with decreasing $\Delta y_{corr}$. The red solid symbols, corresponding to $\Delta y_{corr} = \infty$, are actually computed for $\Delta y_{corr}$ = 100, and are consistent with Eq.(4) of ~\cite{Braun-Munzinger:2018yru}, derived for global baryon number conservation.
In the right panel of Fig.~\ref{fCE} we compare our results for different $\Delta y_{corr}$ to the experimental measurements of the second cumulant of net-protons, as reported by the ALICE collaboration~\cite{RustamovQM17}. For this purpose, we use the acceptance fraction $\alpha$, corresponding to each $\Delta \eta$ in the right panel of Fig.~\ref{fCE} (cf.~\cite{RustamovQM17}) and determined the value of $\kappa_{2}(p-\bar{p})/< n_{p} + n_{\bar{p}} >$ from the left panel of Fig.~\ref{fCE}.  Within the experimental uncertainties, the data are best described by global baryon number conservation but are consistent with $\Delta y_{corr} \ge 5.$  Values of  $\Delta y_{corr}$ smaller than 5 lead to results in disagreement with the experimental measurements. Apparently, effects due to local baryon number conservation  are small in second cumulants of net-protons.

Interestingly, predictions using the HIJING~\cite{Gyulassy:1994ew, Deng:2010mv} generator, presented in the right panel of Fig.~\ref{fCE}, clearly underestimate the experimental data. On the other hand, our calculation with $\Delta y_{corr}$ = 2 is consistent with the HIJING results. This implies that the correlations between protons and anti-protons in the rapidity space obtained from HIJING are too strong ranged, not consistent with the experimental results. 
\section{Conclusions}
In summary, we studied the effects of local baryon conservation laws on fluctuations of net-baryon number.  We demonstrated, by systematic comparison of our results with the measurements of ALICE, that the data are consistent with global baryon number conservation and the according small suppression with increasing $\Delta y$. Contributions due to local baryon number conservation are small if present at all in the second cumulants of net-protons. This is not unexpected as baryon number is distributed, for nuclear collisions at very high energy, over a wide rapidity range in very early hard collisions. A detailed account of this for both baryon and strangeness number is in preparation.

\section*{Acknowledgments}
This work is part of and supported by the DFG Collaborative Research
Centre "SFB 1225 (ISOQUANT)". 

\bibliography{bibliography}

\end{document}